\documentclass{pramana}
\usepackage{multicol}
\usepackage[colorlinks=true,linkcolor=blue,citecolor=blue,urlcolor=blue]{hyperref}
\usepackage{graphicx,amsmath,amssymb,bm,cite}

\newcommand{\be}{\begin{equation}}
\newcommand{\ee}{\end{equation}}
\newcommand{\bea}{\begin{eqnarray}}
\newcommand{\eea}{\end{eqnarray}}
\newcommand{\ba}[1]{\begin{array}{#1}}
\newcommand{\ea}{\end{array}}

\begin{document}

\title{Effect of magnetic field on jet transport coefficient $\hat{q}$}

\author{Debjani Banerjee\textsuperscript{1,*}, Prottoy Das\textsuperscript{1}, Souvik Paul\textsuperscript{2}, Abhi Modak\textsuperscript{1}, Ankita Budhraja\textsuperscript{3}, Sabyasachi Ghosh\textsuperscript{4} \and Sidharth K. Prasad\textsuperscript{1}}

\affilOne{\textsuperscript{1}Department of Physics, Bose Institute, EN 80, Sector-V, Salt Lake, Kolkata 700091, West Bengal, India\\}
\affilTwo{\textsuperscript{2}Department of Physical Sciences, Indian Institute of Science Education and Research, Kolkata, Mohanpur 741246, West Bengal, India\\}
\affilThree{\textsuperscript{3}Tata Institute of Fundamental Research, Homi Bhabha Road, Colaba, Mumbai 400005, Maharashtra, India\\}
\affilFour{\textsuperscript{4}Indian Institute of Technology Bhilai, GEC Campus, Sejbahar, Raipur 492015, Chhattisgarh, India\\}
\date{}

\twocolumn[{

    \maketitle

    \corres{debjanibanerjee771@gmail.com}

    \msinfo{7 January 2023}{31 May 2023}{17 July 2023}

    \begin{abstract}
      We report the estimation of jet transport coefficient, $\hat{q}$ for quark- and gluon-initiated jets using a simple quasi-particle model in absence and presence of magnetic field. This model introduces a temperature and magnetic field-dependent degeneracy factor of partons, which is tuned by fitting the entropy density of lattice quantum chromodynamics data. At a finite magnetic field, $\hat{q}$ for quark jets splits into parallel and perpendicular components whose magnetic field dependence comes from two sources: the field-dependent degeneracy factor and the phase space part guided from the shear viscosity to entropy density ratio. Due to the electrically neutral nature of gluons, the estimation of $\hat{q}$ for gluon jets is affected only by the field-dependent degeneracy factor. In presence of a finite magnetic field, we find a significant enhancement in $\hat{q}$ for both quark- and gluon-initiated jets at low temperature, which gradually decreases towards high temperature.  We compare the obtained results with the earlier calculations based on the anti-de Sitter/conformal field theory correspondence, and a qualitatively similar trend is observed.  The change in $\hat{q}$ in presence of magnetic field is, however, quantitatively different for quark- and gluon-initiated jets. This is an interesting observation which can be explored experimentally to verify the effect of magnetic field on $\hat{q}$.\\
 
    \end{abstract}



}]

\doinum{}
\artcitid{\#\#\#\#}
\volnum{123}
\year{2023}
\pgrange{1--14}
\setcounter{page}{1}
\lp{14}
\sloppy

\section{Introduction}
In high energy hadron-hadron, hadron-nucleus and nucleus-nucleus collisions, the partons produced in large-momentum transfer (large $Q^{2}$) processes undergo multiple splittings (forming shower of partons) and hadronization, leading to a collimated spray of energetic particles known as jets~\cite{JetReview}. Jets are produced very early in the collision due to their small formation time~\cite{FormaTime, formationtime1}. In nucleus-nucleus (AA) collisions, where a thermalized (expanding) medium, quark gluon plasma (QGP) is produced, these jets lose their energy via elastic and inelastic interactions with the medium partons while passing through it, resulting in the suppression of high energy jets in AA collisions relative to that in proton-proton (pp) and proton-nucleus (pA) collisions. This phenomenon of jet energy loss is known as jet quenching~\cite{jq1,jq2,jq3,jet_col} and can be quantified experimentally by a well-known observable, the nuclear modification factor $R_{\rm AA}$, defined as the ratio of high momentum inclusive hadron or jet yields in AA collisions to that in $N_{coll}$ (number of binary collisions) scaled pp collisions. Suppression of high-momentum hadrons and jets was previously measured by NA49, PHENIX, CMS, ATLAS and ALICE experiments at SPS, RHIC and LHC~\cite{NA49RAA, Phenixhsup, CMSptsup, ATLASRaa1, ATLASRaa3, AliceRaa}. Within the framework of theoretical models, jet quenching can be studied to extract and provide constraints on the jet transport coefficient $\hat{q}$ of the QGP, which is defined by the mean square of the momentum transfer between the propagating hard jet and the soft medium per unit path length. The jet initiated by a quark (gluon) is known as quark (gluon) jet. The quark and gluon jets have different sensitivity to the medium due to differences in their color degrees of freedom. The amount of transverse momentum broadening of the jet can be related to density of the gluon distribution in the medium~\cite{q-hat1,q-hat2}. This allows a temperature-dependent proportional relation between $\hat{q}$ and gluon density.
\par
The phenomena of jet quenching has been very well explored by various theoretical models~\cite{jet_col,Feal1,Chen,Chen2,Mustafa,Martini-model,Liu,Andres,Xie,Xie2,Han}, viz. GLV-CUJET~\cite{GLV-1,GLV-2,CUJET}, MARTINI~\cite{Martini-model}, MCGILL-AMY~\cite{Mustafa}, HT-M~\cite{jq3,HT-M-2}, HT-BW~\cite{HT-BW-1,HT-BW-2,Chen}, JEWEL~\cite{JEWEL1st, JEWEL2nd, JEWEL3rd,jetRAAALICE} etc. The value of $\hat{q}$ is estimated in some of these models by explaining the experimentally measured quantity $R_{\rm AA}$. To encapsulate the parton energy loss, different approaches are used by these models. GLV-CUJET relies on multiple scattering in the medium for energy loss and is controlled by the strong coupling constant, the Debye screening mass and the density of scattering centers. HT-BW, HT-M use high-twist (HT) approach where the energy loss is only affected by $\hat{q}$. Within the hard-thermal-loop (HTL) resummed thermal field theory-based MARTINI and MCGILL-AMY models, the only controlling parameter for energy loss is the strong coupling constant. Based on $R_{\rm AA}$ for neutral pion spectra reported by PHENIX experiment~\cite{PHENIX1,PHENIX2}, the extracted values of the jet transport coefficient at the initial time of QGP formation, $\hat{q}_{0}$, are 0.9 $_{-0.04}^{+0.05}$ GeV$^2$/fm~\cite{Chen} for 0--10\% central and 1.2 $\pm$ 0.3 GeV$^2$/fm~\cite{jet_col} for 0--5\% central Au--Au collisions at $\sqrt{s_{\rm NN}}$ = 0.2 TeV for $\tau_{0}$ = 0.6 fm/$c$. Similarly based on combined ALICE~\cite{ALICE} and CMS~\cite{CMS} data on charged hadron spectra in 0--5\% central Pb--Pb collisions at $\sqrt{s_{\rm NN}}$ = 2.76 TeV, the extracted value of $\hat{q}_{0}$ at $\tau_{0}$ = 0.6 fm/$c$ is 2.2 $\pm$ 0.5 GeV$^2$/fm~\cite{jet_col}. However, recently using the combined results of both  $R_{\rm AA}$ and  $I_{\rm AA}$ for Au--Au collisions at $\sqrt{s_{\rm NN}}$ = 0.2 TeV~\cite{PHENIX1,PHENIX2,STARNeutral1,STARNeutral2}, Pb--Pb collisions at $\sqrt{s_{\rm NN}}$ = 2.76 TeV~\cite{CMS,ALICE,ATLASCharged,ALICEDiHadron,CMSDiHadron,ALICENeutral} and 5.02 TeV~\cite{CMSCharged5TeV,ALICECharged5TeV}, including 0--50\% central events, it has been shown that $\hat{q}/T^{3}$ decreases with temperature from 5 $\pm$ 1 near the critical temperature $T_{\rm c}$ to 1.1 $\pm$ 0.3 at 3$T_{\rm c}$~\cite{qhatNew}. 
\par
In non-central nucleus-nucleus collisions, a large magnetic field ($B$) is expected to be created due to moving charges (spectators) at relativistic energies. The strength of this magnetic field produced immediately after the collision is predicted to reach up to a value of $m_{\pi}^{2}$ ($\sim 10^{18}$ G) at RHIC and 10$m_{\pi}^{2}$ at LHC~\cite{Tuchin_B}. It is very important to study the effect of this large magnetic field on various QGP properties such as jet transport coefficient, $\hat{q}$.
\par 
This article explores the effect of magnetic field on $\hat{q}$. Without going into the microscopic calculation, a prescription similar to Ref.~\cite{Chen} is followed where the transition from $\hat{q}_{0}$ to $\hat{q}_{N}$ (jet transport coefficient at the centre of the cold nuclear matter in a large nucleus) and vice-versa are mapped through a quasi-particle type description~\cite{SS_JPG}, based on the thermodynamics of lattice quantum chromodynamics (LQCD)~\cite{LQCD1_B0,LQCD2_B0}, to estimate $\hat{q}$ in absence of magnetic field. Later it is extended for a finite magnetic field picture, based on the LQCD magneto-thermodynamical quantities~\cite{Bali1,Bali2}. Similar kind of magnetic field extension of quasi-particle models has been addressed in Refs.~\cite{PB1_19,PB2_Tawfik,PB3_Tawfik,PB4_Farias}, based on effective QCD models. This model provides a simple and easy-dealing tool to map LQCD magneto-thermodynamical quantities. To add an anisotropic phase space part, at finite temperature $T$ and magnetic field $B$, the relation between shear viscosity $\eta$ and $\hat{q}$ in presence of magnetic field~\cite{Majumder_PRL,XNWang1,Gyulassy2,Gyulassy3,Shuryak,Ayala} is considered. Similar kind of connection between shear viscosity and $\hat{q}$ in absence of magnetic field was studied in Ref.~\cite{aditya}.
\par
This article is organized as follows. Section~\ref{mapping} describes the mapping of LQCD thermodynamical quantities in presence of magnetic field by introducing a temperature and magnetic field-dependent degeneracy factor. The formalism to estimate jet transport coefficient using quasi-particle description, the magnetic field-dependent jet transport coefficient using the temperature and magnetic field-dependent phase space part of shear viscosity are discussed in Sec.~\ref{qhat}. The results obtained in this work and their comparisons with the anti-de Sitter/conformal field theory (AdS/CFT) correspondence~\cite{Li,Liu:2006ug} are presented in Sec.~\ref{result}. The outcome of this work is summarized in Sec.~\ref{sum}.

\begin{figure*}[h!]
    \centering
    \includegraphics[width=0.48\textwidth]{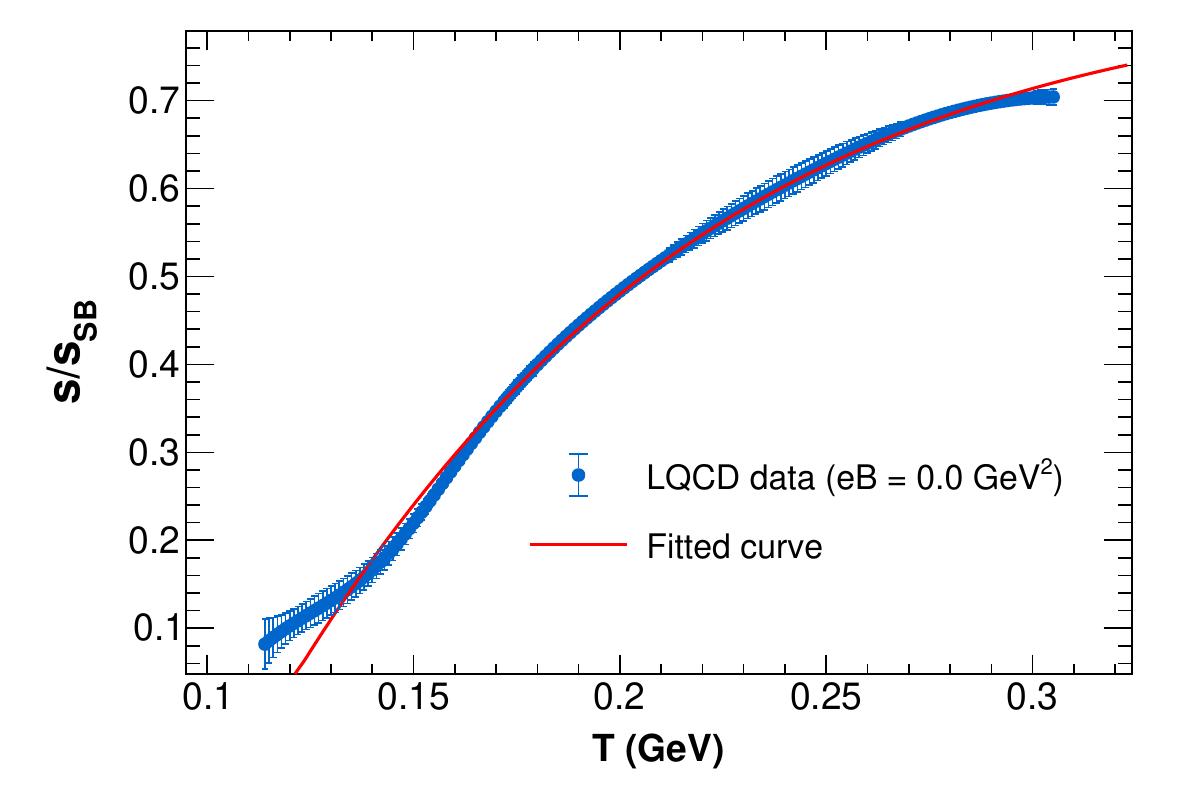}
    \includegraphics[width=0.48\textwidth]{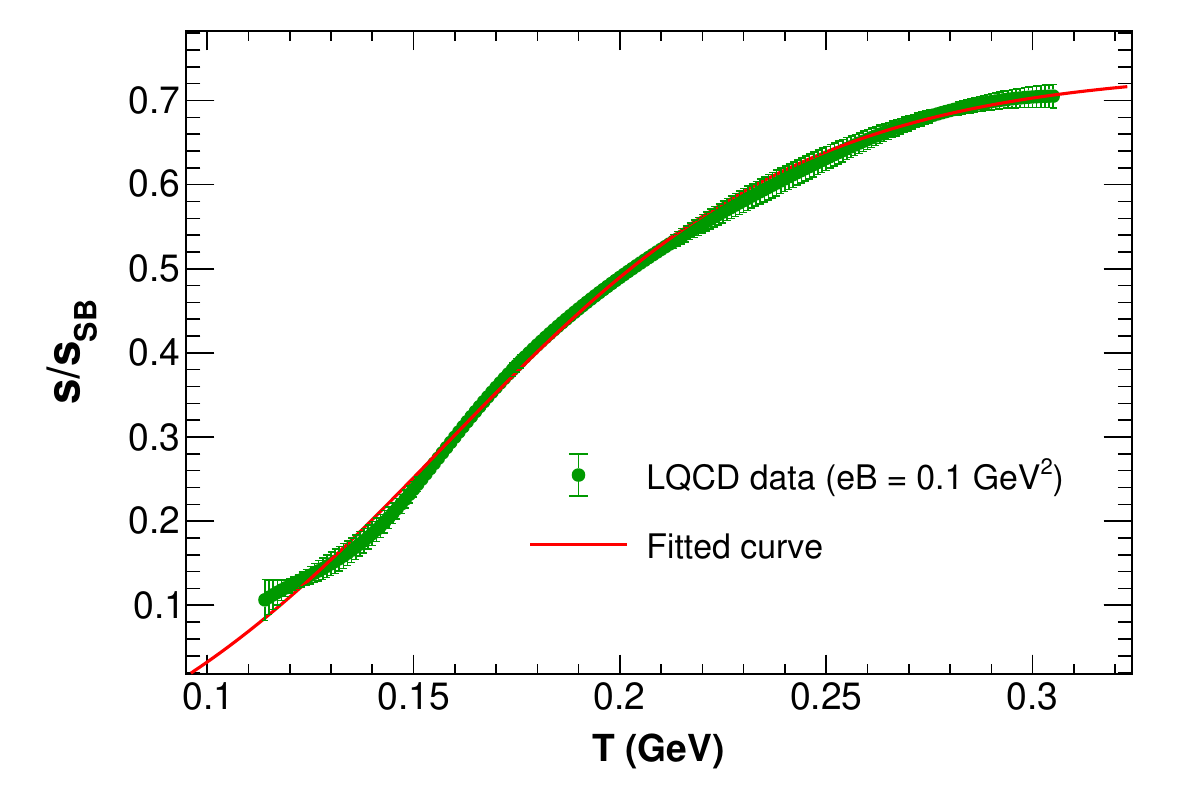}
    \includegraphics[width=0.48\textwidth]{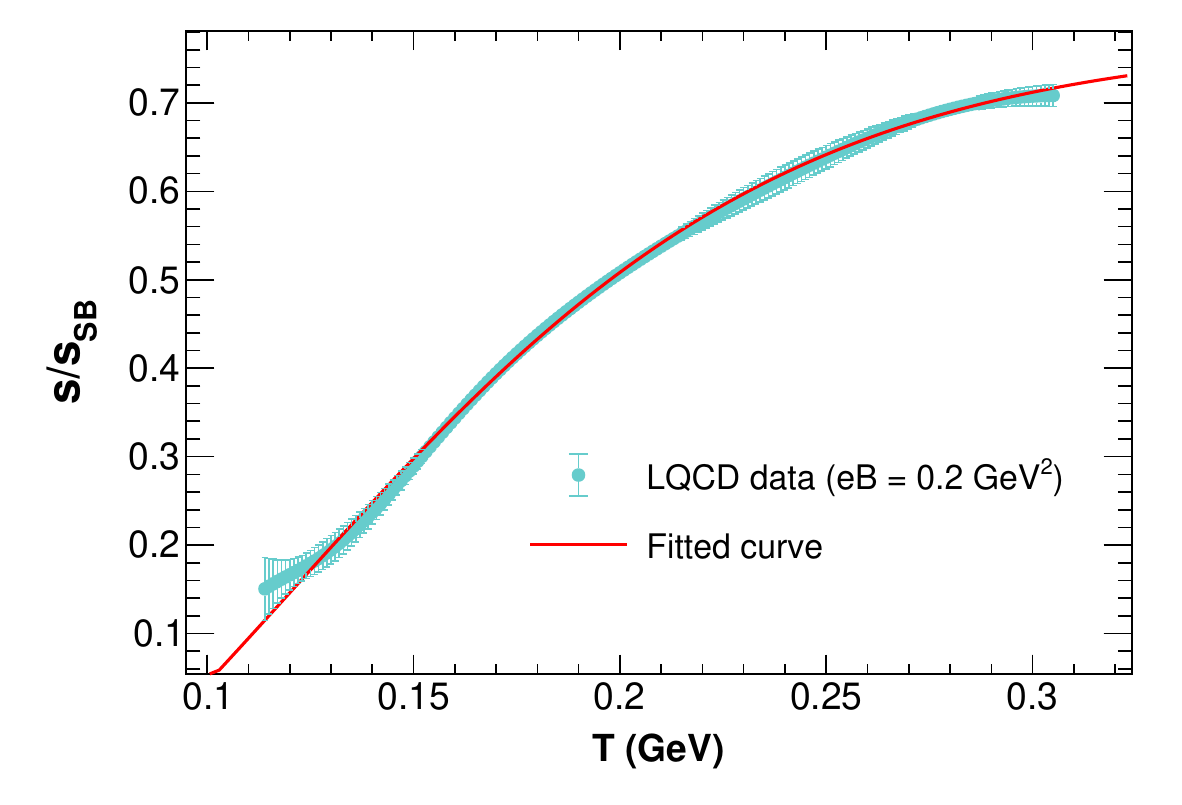}
    \includegraphics[width=0.48\textwidth]{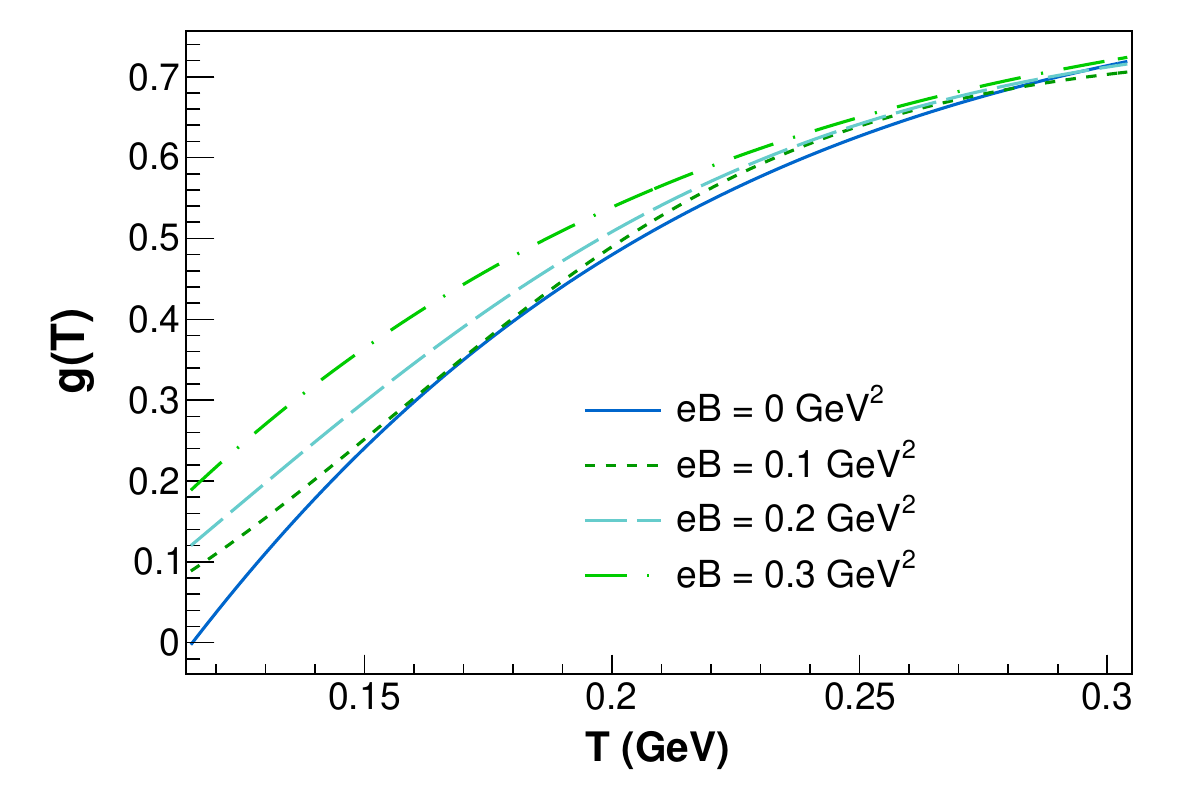} 
    \caption{Top left, top right and bottom left: LQCD data points (with error bars)~\cite{Bali2} and fitted curves using Eq.~(\ref{g_a0123}) (solid line) of normalized entropy density, normalized by its SB limit ($s_{SB}$)  vs $T$ at $eB=0$ (blue line), $eB=0.1$ GeV$^2$ (green line) and $eB=0.2$ GeV$^2$ (cyan line). Bottom right: Corresponding $T$-dependent fraction $g(T)$ [Eq.~(\ref{g_a0123})] at different $eB$ (to be multiplied with total degeneracy factor).}
  \label{gvsT}
\end{figure*}

\section{Parametrization of LQCD magneto-thermodynamical data using quasi-particle model}
\label{mapping}
This section describes the procedure adopted to map LQCD thermodynamical data~\cite{Bali2} in presence of magnetic field following the steps outlined in~\cite{SS_JPG}. In terms of the Fermi-Dirac (FD) distribution function of quarks and the Bose-Einstein (BE) distribution function of gluons, the energy density ($\epsilon$) and pressure ($P$) of the QGP system can be expressed as
\begin{align}
  \epsilon_{QGP} ={}& \frac{g_{g}}{(2\pi)^{3}} \int_{0}^{\infty} \frac{\omega_{g}}{e^{\beta\omega_{g}} - 1} {d^{3}\vec{k}}\nonumber\\
  & + \frac{g_{Q} }{(2\pi)^{3}}\int_{0}^{\infty} \frac{\omega_{Q}}{e^{\beta\omega_{Q}} + 1} {d^{3}\vec{k}}
  \label{e_QGP}
\end{align}
and
\begin{align}
  P_{QGP} ={}&  \frac{g_{g}}{3(2\pi)^{3}} \int_{0}^{\infty} \Big(\frac{\vec{k}^{2}}{\omega_{g}}\Big) \frac{1}{e^{\beta\omega_{g}} - 1} {d^{3}\vec{k}}\nonumber\\
  & + \frac{g_{Q} }{3(2\pi)^{3}}\int_{0}^{\infty} \Big(\frac{\vec{k}^{2}}{\omega_{Q}}\Big) \frac{1}{e^{\beta\omega_{Q}} + 1} {d^{3}\vec{k}}
  \label{P_QGP}
\end{align}
where $\omega_{Q}$, $\omega_g$ are energies and
\begin{align}
  g_{Q} &= (\rm spin) \times (\rm particle/anti particle) \times (\rm color) \times (\rm flavor) \nonumber\\
  &= 2 \times 2 \times 3 \times 3 = 36, \nonumber \\
  g_{g} &= (\rm spin) \times (\rm color) = 2 \times 8 = 16,
  \label{degen_QG}
\end{align}
are degeneracy factors of quarks and gluons respectively. At zero net quark density or chemical potential, the entropy density ($s$) can be expressed in terms of $P$ and $\epsilon$~\cite{JD2} as
\begin{equation}
  s_{QGP} = \frac{P_{QGP}+\epsilon_{QGP}}{T}~.
  \label{s_QGP}
\end{equation}
For a massless QGP system (derivations are shown in Appendix A),
\begin{align}
P_{QGP}&=\Big[g_g+g_{Q}\Big(\frac{7}{8}\Big)\Big]\frac{\pi^2}{90}T^4\approx 5.2~T^4~,
\nonumber \\
\epsilon_{QGP}&= \Big[g_g+g_{Q}\Big(\frac{7}{8}\Big)\Big]\frac{3\pi^2}{90}T^4\approx 15.6~T^4~,
\nonumber \\
s_{QGP}&=\Big[g_g+g_{Q}\Big(\frac{7}{8}\Big)\Big]\frac{4\pi^2}{90}T^3\approx 20.8~T^3~.
\label{SB_QGP}
\end{align}
Now, if one sees the LQCD data of $P(T)$, $\epsilon(T)$, $s(T)$ at $eB=0$ from Ref.~\cite{Bali2}, then one can notice that the data points always remain lower than their massless limits, i.e. $P/T^4$ $<$ 5.2, $\epsilon/T^4$ $<$ 15.6, $s/T^4$ $<$ 20.8. A rich QCD interaction in the non-perturbative domain is responsible for this suppression, and it is LQCD calculation that provides useful insights in this domain, where pQCD\footnote{Though pQCD estimation remains little lower than massless or SB limits but not too much lower as LQCD estimations. So for simplicity, we may consider massless and pQCD as equivalent estimations.} 
does not work well. One can map this temperature-dependent suppression of $P(T)$, $\epsilon(T)$, $s(T)$~\cite{SS_JPG} by imposing a temperature-dependent fraction $g(T)$, multiplied with the total degeneracy factor of QGP. So, this picture assumes that while going from high temperature to low temperature, the degeneracy factor of the massless QGP system is gradually reduced. Instead of considering massless QGP, one can consider physical mass of quark in Eqs.~(\ref{e_QGP}), (\ref{P_QGP}) and (\ref{s_QGP}) and then fit their $P$, $\epsilon$ and $s$ with LQCD data by tuning $g(T)$. However, in this work, the massless QGP expression is adopted due to its visibility for simple analytic structure. One should also note that the numerical difference between massless QGP and QGP with physical mass is quite small.
\par
Now, let us proceed with a finite magnetic field picture. The presence of finite magnetic field results in an anisotropy in the pressure. However, pressure parallel to magnetic field will follow the simple thermodynamic relation
\begin{equation}
  s=\frac{\epsilon +P_{\parallel}}{T}
  \label{s_ep_p}
\end{equation}
remains the same~\cite{Bali2}. Therefore, the earlier expressions can still be used for finite $B$ picture by introducing a $T$, $B$-dependent parameter $g(T,B)$. The parametric form of $g(T,B)$ can be obtained by mapping LQCD data of $P(T,B)$, $\epsilon(T,B)$, $s(T,B)$~\cite{Bali1,Bali2}:
\begin{equation}
  g(T,B)= a_0 - \frac{a_1}{e^{a_2(T-0.17)}+ a_3}~.
  \label{g_a0123}
\end{equation}

In this work,  the LQCD data of the entropy density $s(T,B)$ are mapped using Eq.~(\ref{g_a0123}) to obatin the fitting parameters $a_{0,1,2,3}$  of the parametric form of $g(T,B)$ because one of the aims is also to estimate $s/\eta$ later, which is connected with the dimensionless ratio $\hat{q}/T^3$. Figure~\ref{gvsT} shows the LQCD data points and their fitted curves of normalized entropy density at $eB$ = 0, 0.1 and 0.2 GeV$^2$ using Eq.~(\ref{g_a0123}) and $T$-dependent fraction $g(T)$ at different $eB$, respectively. The values corresponding to the tuning parameters ($a_{0,1,2,3}$) are summarized in Table~\ref{tab:table1}.
\begin{table}[h!]
  \begin{center}
    \caption{Different values of $a_{0,1,2,3}$ given in Eq.~(\ref{g_a0123}) for different magnetic field strengths.\\}
    \label{tab:table1}
    \begin{tabular}{ |c|c|c|c|c| } 
      \hline
      $\mathbf{\textit{eB}(GeV^2)}$ & $\mathbf{a_0}$ & $\mathbf{a_1}$ & $\mathbf{a_2}$ & $\mathbf{a_3}$\\ 
      \hline
      0.0 & -0.26 & -1.49 & -21.58 & 1.49\\ 
      0.1 & -0.16 & -1.20 & -22.61 & 1.33\\ 
      0.2 & -0.47 & -2.82 & -16.51 & 2.26\\
      \hline
    \end{tabular}
  \end{center}
  \label{tab1}
\end{table}

\section{Jet transport coefficient $\hat{q}$}
\label{qhat}
The temperature-dependent jet transport coefficient $\hat{q}(T)$ of a medium of hadron resonance gas can be approximated, following Ref.~\cite{Chen}, as:
\begin{equation}
  \hat{q} (T) = \frac{\hat{q_{N}}}{\rho_{N}} \rho_{h} (T)~,
  \label{qh_rho}
\end{equation}
where $\hat{q}_N\approx 0.02$ GeV$^2$/fm~\cite{qN, q_N1} is the jet transport coefficient at the center of the cold nuclear matter in a large nucleus, $\rho_N=0.17$ fm$^{-3}$ is nuclear saturation density~\cite{NSD} and $\rho_h$ is hadronic matter density. In Ref.~\cite{Chen}, $\rho_h(T)$ is obtained using hadron resonance gas (HRG) model within the hadronic temperature range. The thermodynamics in the HRG model is found to match well with LQCD data in the hadronic temperature range. Here, a quasi-particle model with temperature-dependent degeneracy factor, which is obtained from parameterizing LQCD data as discussed in Sec.~\ref{mapping}, has been used as an alternative approach for estimations of thermodynamical quantities and it is not only used for hadronic temperature domain but also for quark temperature domain, beyond the quark-hadron phase transition. Because the jet transport coefficient is expected to be proportional to the effective density of the scatterers in the medium, which is dominated by gluons~\cite{q-hat1,XNWang1,Chen}, the hadronic matter density $\rho_h(T)$ can be replaced by the density of gluons in the medium $\rho_{G}(T)$ and consider it for the entire temperature range.
\par
Similar to Eqs.~(\ref{e_QGP}) and (\ref{P_QGP}), the gluon density $\rho_{G}$ can be expressed as:
\begin{equation}
  \rho_{G}= \frac{g_g}{(2\pi)^3}\int_{0}^{\infty}\frac {1}{e^{\beta \omega_g}-1} {d^3\vec k}.
  \label{rho_G}
\end{equation}
Massless gluons provide analytic expression~\cite{JD2}:
\begin{align}
  \rho_{G}&= \Big[g_g\frac{\zeta(3)}{\pi^2}\Big]T^3\nonumber\\
  &= 1.94~T^3~,
\end{align}
which can be considered as high temperature~($T$ $\rightarrow$ $\infty$)\footnote{Infinity means large temperature (say $T\geq 0.3$ GeV).} limiting values, where pQCD\footnote{Here for simplicity, we are assuming that massless non-interacting limit or SB limit of gluon density are approximately equal to its pQCD results.} works well. Therefore, grossly one can consider it as gluon density in pQCD domain and its corresponding high temperature limiting expression of $\hat{q}$ will be  
\begin{equation}
  \hat{q_{\infty}} = \frac{\hat{q_{N}}}{\rho_{N}} \times 1.94~T^3 = 3.03 \times 1.94~T^3
\end{equation}
whose normalized value $\hat{q}_{\infty}/T^3$ saturates at 5.87. 
\par
In the non-perturbative QCD domain, one can make a rough estimation of the gluon density by multiplying 1.94~$T^3$ with the $T$- and $B$-dependent degeneracy factor $g(T,B)$. Following Eq.~(\ref{qh_rho}) and taking $\hat{q}_N$ as the reference point, the jet transport coefficient $\hat{q}(T,B)$ for the entire temperature range can be written as:
\begin{align}
  \hat{q}(T,B) &= \frac{\hat{q}_N}{\rho_N}\times g(T,B)\times g_g\frac{\zeta(3)}{\pi^2}~T^3 \nonumber\\
  &= 3.03\times\Big[a_0 - \frac{a_1}{e^{a_2(T-0.17)}+ a_3}\Big]\times 1.94~T^3
  \label{qh_rho_qN}
\end{align}
\par
The above expression is built from the cold nuclear matter reference point and it extends from hadronic to quark temperature domain. As an alternative way, one can make initial value of jet quenching $\hat{q}_{0}$ as reference point and extend it from high (quark) to low (hadronic) temperature range. In this case, assuming an equivalence between $\hat{q}_0$ and $\hat{q}$ one can normalize Eq.~(\ref{qh_rho_qN}) by multiplying with $\hat{q}_{0}/\hat{q}_{\infty}$:
\begin{align}
  \hat{q}(T,B)&= \frac{\hat{q}_N}{\rho_N}\times g(T,B)\times g_g\frac{\zeta(3)}{\pi^2}~T^3\Big(\frac{\hat{q}_0}{\hat{q}_{\infty}}\Big)\nonumber\\
  &= g(T,B)\times \hat{q}_0\nonumber\\
  &= \Big[a_0 - \frac{a_1}{e^{a_2(T-0.17)}+ a_3}\Big]\times 0.9 ~{\rm GeV}^2/{\rm fm}.
  \label{qh_rho_q0}
\end{align}
Therefore, either by using Eq.~(\ref{qh_rho_qN}) or Eq.~(\ref{qh_rho_q0}), one can make an estimation of $\hat{q}(T,B)$ in the quasi-particle model. The parameters ($a_0$, $a_1$, $a_2$ and $a_3$) of the temperature and magnetic field-dependent degeneracy factor $g(T,B)$ in the model are obtained by fitting the LQCD data, as discussed in Sec.~\ref{mapping}. For the estimation of $\hat{q}(T)$ in absence of magnetic field, the model uses LQCD data at e$B$ = 0 to obtain its fit parameters. At finite magnetic field, the parameters of $g(T,B)$ are obtained by fitting the LQCD magneto-thermodynamical data at $eB$ = 0.2 GeV$^2$. This extension to finite magnetic field domain has different impacts in the estimation of $\hat{q}(T, B)$ for quark and gluon jets. In case of gluon jets, only the medium gets influenced by the presence of magnetic field, whereas quark jets being electrically charged also experience the effect of magnetic field along with the medium. Similar to other transport coefficients such as shear viscosity, electrical conductivity of QGP, $\hat{q}(T,B)$ can have a multicomponent structure (e.g. $\hat{q}_{\parallel},~\hat{q}_{\perp}$, etc.) for quark jets which are discussed in the next subsection.
\subsection{Phase space component of $\hat{q}(T,B)$ for quark jets}
\label{Pspace}
According to Refs.~\cite{Majumder_PRL,XNWang1,Gyulassy2,Gyulassy3,Shuryak}, the jet transport coefficient is connected with the shear viscosity coefficient, $\eta$. Here, $\hat{q}(T,B)$ is calculated from the knowledge of $\eta(T,B)$ profile. The advantage of this approach lies in the fact that the $T$, $B$-dependent phase space part of $\eta(T,B)$ is well studied in Refs.~\cite{Landau,XGH1,XGH2,Tuchin,Li_shear,Asutosh,G_NJL_B,Nam,JD2,HRGB,HM3,Denicol,BAMPS,JD_eta}. So, based on that knowledge, the $T$, $B$-dependent phase space part has been invoked into $\hat{q}(T,B)$.
\par
Let us take a quick revisit of shear viscosity expressions at finite temperature and magnetic field. First, one can consider the case of zero magnetic field and then move to the non-zero magnetic field picture. According to the macroscopic fluid definition, shear viscosity $\eta$ is the proportionality constant between viscous stress tensor $\pi^{ij}$ and velocity gradient tensor $C^{ij}$~\cite{JD_eta}, i.e.
\begin{equation}
\pi^{ij}=\eta \, C^{ij}~,
\label{Macro_B0}
\end{equation}
which is the relativistic and tensor form of the so-called Newton's law of viscosity. In the microscopic kinetic theory approach, the viscous stress tensor can be connected with the deviation (from the equilibrium distribution function $f_0$) $\delta f=Ck^nk^lC_{kl}\beta f_0(1\mp f_0)$ as
\begin{align}
  \pi^{ij} &= g\int \frac{d^3\vec k}{(2\pi)^3}\frac{k^ik^j}{\omega}\delta f \nonumber\\
  &= g\int \frac{d^3\vec k}{(2\pi)^3}\frac{k^ik^j}{\omega}Ck^nk^lC_{kl}\beta f_0(1- a f_0)~,
  \label{Micro_B0}
\end{align}
where $f_0=1/[e^{\beta\omega}+a]$ denotes the Fermi-Dirac (FD) and Bose-Einstein (BE) distribution functions for $a=\pm 1$ respectively and $\omega=\sqrt{\vec k^2+m^2}$ is the energy of medium constituent with mass $m$ and degeneracy factor $g$. Connecting the macroscopic Eq.~(\ref{Macro_B0}) and microscopic Eq.~(\ref{Micro_B0}), one can get the shear viscosity tensor:
\begin{equation}
  \eta^{ijkl} = g\int \frac{d^3\vec k}{(2\pi)^3}\frac{k^ik^jk^nk^l}{\omega^2}\tau_c\beta f_0(1- a f_0)~,
  \label{sh_ijkl}
\end{equation}
whose isotropic expression will be
\begin{equation}
  \eta = \frac{g}{15}\int \frac{d^3\vec k}{(2\pi)^3}\frac{\vec k^4}{\omega ^2}\tau_c\beta f_0(1- a f_0)~.
\end{equation}
The unknown constant $C$ in Eq.~(\ref{Micro_B0}) was obtained in terms of relaxation time $\tau_c$ with the help of relaxation time approximation (RTA) based relativistic Boltzmann equation (RBE)~\cite{JD_eta}.
\par
In presence of magnetic field, five independent traceless tensors are proposed in Refs.~\cite{Landau, XGH1, XGH2} instead of a single traceless velocity gradient tensor $C^{ij}$. Two interconnected sets of five shear viscosity components ${\tilde\eta}_{0,1,2,3,4}$~\cite{JD_eta,Asutosh,HRGB} and $\eta_{0,1,2,3,4}$~\cite{Denicol,JD_eta} are obtained.  The two main components based on the direction of the applied magnetic field are as follows:
\begin{align}
  \frac{\eta_{\parallel}}{\eta} &= \frac{{\tilde\eta}_2}{\eta} = \frac{(\eta_0 + \eta_2)}{\eta} = \frac{1}{1+(\tau_c/\tau_B)^2}\nonumber\\
  \frac{\eta_{\perp}}{\eta} &= \frac{{\tilde\eta}_1}{\eta} = \frac{\eta_0}{\eta} = \frac{1}{1+4(\tau_c/\tau_B)^2}~,
  \label{eta_perPar}
\end{align}
where another time scale $\tau_B=E_{jet}/e_{q}B$ ($E_{jet}$ is the jet energy) enters into the picture along with relaxation time $\tau_c$. Here, $e_{q}$ denotes the electric charge of quarks. It is important to note that the definition of $\tau_B$ is only valid for quarks and not gluons due to their chargeless nature. A simplified general expression of shear viscosity for massless case is as follows, 
\begin{align}
  \eta &= \frac{4g\:\tau_c}{5\pi^2}\zeta(4) \: T^4 ~~{\rm for~BE}\nonumber\\
  &= \Big(\frac{7}{8}\Big)\frac{4g\:\tau_c}{5\pi^2}\zeta(4) \: T^4 ~~{\rm for~FD}~.
  \label{eta_anl}
\end{align}
Now, for massless 3 flavor QGP at $B=0$, one can get
\begin{align}
  \eta &= \Big[16+\frac{7}{8}36\Big]\frac{4\:\tau_c}{5\pi^2}\zeta(4) \: T^4~, \nonumber\\
  s &= \Big[16+\frac{7}{8}36\Big]\frac{4\:\tau_c}{\pi^2}\zeta(4) \: T^3~, \nonumber\\
  \Rightarrow\frac{\eta}{s} &= \frac{\tau_c T}{5}~.
  \label{eta_s}
\end{align}
For $B\neq 0$, one gets
\begin{align}
  \frac{\eta_{\parallel}}{s} &= \frac{1}{\Big[16+\frac{7}{8}36\Big]}\Big[16+\frac{7}{8}12\sum_{u,d,s}\frac{1}{1+(\tau_c/\tau_B)^2}\Big]\frac{\tau_c T}{5} \nonumber\\
  \frac{\eta_{\perp}}{s} &= \frac{1}{\Big[16+\frac{7}{8}36\Big]}\Big[16+\frac{7}{8}12\sum_{u,d,s}\frac{1}{1+4(\tau_c/\tau_B)^2}\Big]\frac{\tau_c T}{5}~.
  \label{eta_s_par}
\end{align}
Now, by roughly connecting $\hat{q}\propto s/\eta$ at $B=0$ and $\hat{q}_{\parallel,\perp}\propto s/\eta_{\parallel,\perp}$ at $B\neq 0$ one may write:
\begin{align}
  \frac{\hat{q}_{\parallel}(B)}{\hat{q}(B=0)}&= \frac{s/\eta_{\parallel}}{s/\eta} = \frac{47.5}{\Big[16+\frac{7}{8}12\sum\limits_{u,d,s}\frac{1}{1+(\tau_c/\tau_B)^2}\Big]} \nonumber\\
  \frac{\hat{q}_{\perp}(B)}{\hat{q}(B=0)}&= \frac{s/\eta_{\perp}}{s/\eta} = \frac{47.5}{\Big[16+\frac{7}{8}12\sum\limits_{u,d,s}\frac{1}{1+4(\tau_c/\tau_B)^2}\Big]}~.
  \label{qh_q0}
\end{align}
Now realizing $\hat{q}(B=0)$ in terms of a simple quasi-particle form either $\hat{q}(B=0) = g(T) \times 5.87~T^3$ from Eq.~(\ref{qh_rho_qN}) or $\hat{q}(B=0)=g(T)\times \hat{q}_0$ from Eq.~(\ref{qh_rho_q0}), $\hat{q}_{\parallel,\perp}(T,B)$ can be expressed as:

\begin{align}
  \hat{q}_{\parallel}(T,B) &=  \frac{47.5}{\Big[16+\frac{7}{8}12\sum\limits_{u,d,s}\frac{1}{1+(\tau_c/\tau_B)^2}\Big]} \times \Big[g(T)\times 5.87~T^3\Big]\nonumber\\
  \text{or~~~~~~~~~} \nonumber\\
  \hat{q}_{\parallel}(T,B) &= \frac{47.5}{\Big[16+\frac{7}{8}12\sum\limits_{u,d,s}\frac{1}{1+(\tau_c/\tau_B)^2}\Big]} \times \Big[g(T)\hat{q}_0\Big]
  \label{q_par}
\end{align}
\begin{align}
  \hat{q}_{\perp}(T,B) &= \frac{47.5}{\Big[16+\frac{7}{8}12\sum\limits_{u,d,s}\frac{1}{1+4(\tau_c/\tau_B)^2}\Big]} \times \Big[g(T)\times 5.87~T^3\Big] \nonumber\\
  \text{or~~~~~~~~~~} \nonumber\\
  \hat{q}_{\perp}(T,B) &= \frac{47.5}{\Big[16+\frac{7}{8}12\sum\limits_{u,d,s}\frac{1}{1+4(\tau_c/\tau_B)^2}\Big]} \times \Big[g(T)\hat{q}_0\Big]
  \label{q_per}
\end{align}
For more rich $B$-dependent structure, $g(T)$ is replaced by $g(T,B)$. Here, jet transport coefficient for quark jets has parallel and perpendicular components ($\hat{q}^q_{\parallel,\perp}$) which is not expected for gluon jet transport coefficient ($\hat{q}^g$). Therefore, the final expressions of gluon and quark jets at finite $B$ can be written as:
\begin{align}
  \hat{q}^{g}(T,B) &= g(T,B)\times 5.87~T^3 \nonumber\\
  \text{or~~~~~~~~~~} \nonumber\\
  \hat{q}^{g}(T,B) &= g(T,B)\hat{q}_0
  \label{q_gluon}
\end{align}
\begin{align}
  \hat{q}^{q}_{\parallel} (T,B) ={}& \frac{47.5}{\Big[16+\frac{7}{8}12\sum\limits_{u,d,s}\frac{1}{1+(\tau_c/\tau_B)^2}\Big]}\nonumber\\
  & \times \Big[g(T,B)\times 5.87~T^3\Big]\nonumber\\
  \text{or~~~~~~~~~~~~~~} \nonumber\\
  \hat{q}^{q}_{\parallel} (T,B) ={}& \frac{47.5}{\Big[16+\frac{7}{8}12\sum\limits_{u,d,s}\frac{1}{1+(\tau_c/\tau_B)^2}\Big]}\nonumber\\
  & \times \Big[g(T,B)\hat{q}_0\Big]
  \label{q_par_quark}
\end{align}
\begin{align}
  \hat{q}^{q}_{\perp}(T,B) ={}& \frac{47.5}{\Big[16+\frac{7}{8}12\sum\limits_{u,d,s}\frac{1}{1+4(\tau_c/\tau_B)^2}\Big]}\nonumber\\
  & \times \Big[g(T,B)\times 5.87~T^3\Big]\nonumber\\
  \text{or~~~~~~~~~~~~~~}\nonumber \\
  \hat{q}^{q}_{\perp}(T,B) ={}& \frac{47.5}{\Big[16+\frac{7}{8}12\sum\limits_{u,d,s}\frac{1}{1+4(\tau_c/\tau_B)^2}\Big]}\nonumber\\
  & \times \Big[g(T,B)\hat{q}_0\Big]
 \label{q_per_quark}
\end{align}

\section{Results and discussions}
\label{result}
\subsection{$\hat{q}$ in absence of magnetic field}
\begin{figure*}[h!]
  \centering
  \includegraphics[width=0.48\textwidth]{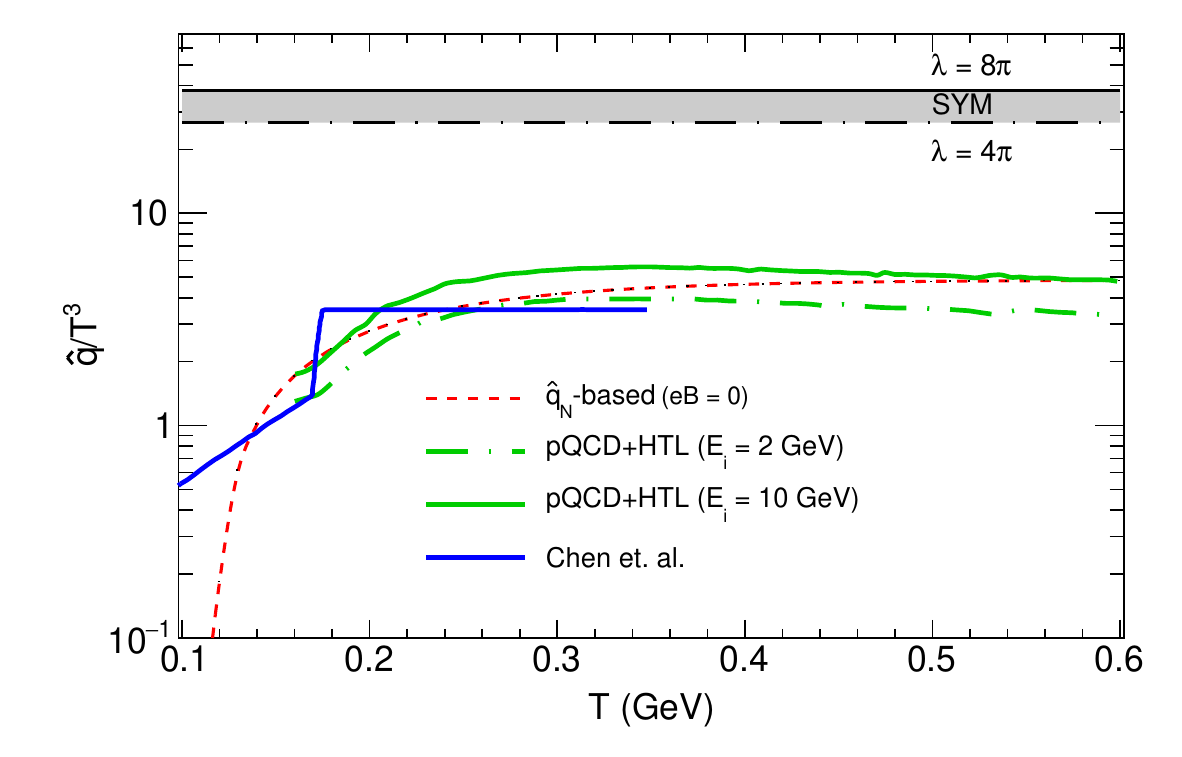}
  \includegraphics[width=0.48\textwidth]{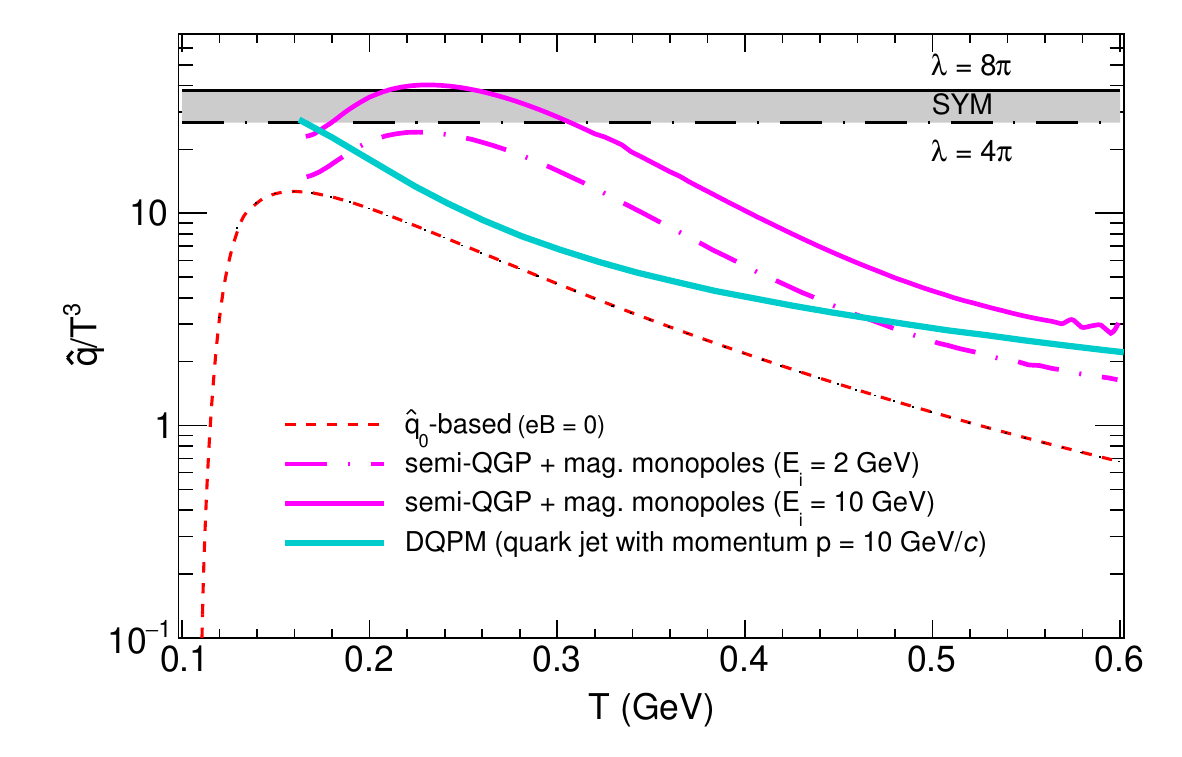}
  \caption{Temperature dependence of $\hat{q}/T^{3}$ based on $\hat{q}_{N}$ and $\hat{q}_{0}$ in absence of magnetic field are depicted by the red dashed line in left and right panels respectively. The green (left) and magenta (right) lines represent pQCD + HTL approximations and semi-QGP + magnetic monopoles estimations respectively for quark jet of initial energy $E_{i}$ = 2 GeV (dash-dotted) and 10 GeV (solid) in absence of magnetic field~\cite{Gyulassy2}. The blue curve (left) is taken from Chen et al., Ref.~\cite{Chen} to compare with $\hat{q}_{N}$ based results. The cyan curve (right) shows DQPM results from Ref.~\cite{DQPM}. Black dash-dotted and solid lines represent $\hat{q}$ of an isotropic $\cal{N}$ = 4 SYM plasma at zero magnetic field for $\lambda$ = 4$\pi$ and 8$\pi$ respectively.}
  \label{qbyT3_all}
\end{figure*}

First, the quasi-particle based numerical estimation of $\hat{q}$ in absence of magnetic field is calibrated with its standard values available in existing references and then its modification in presence of magnetic field is studied.  The temperature dependence of $\hat{q}/T^{3}$ is estimated using the quasi-particle model in absence of magnetic field as shown in Fig.~\ref{qbyT3_all} and compared with earlier estimations of Refs.~\cite{Gyulassy2,pQCDHTL,SemiQGP1,SemiQGP2,SemiQGP3,SemiQGP4,monopole1,monopole2,monopole3}. In Fig.~\ref{qbyT3_all} (left), the red dashed line shows the $\hat{q}$ estimation using Eq.~(\ref{qh_rho_qN}) starting from low (hadronic) to high (QGP) temperature taking $\hat{q}$ at the center of cold nuclear matter ($\hat{q}_N$ $\approx$ 0.02 GeV$^2$/fm) as reference point.
It represents a crossover type transition in $\hat{q}(T)$  due to the continuous profile of $g(T)$, which carries the information of LQCD thermodynamics. The solid blue line in the left panel of Fig.~\ref{qbyT3_all} shows data taken from Ref.~\cite{Chen}. It represents a first-order type transition in $\hat{q}$ owing to the fact that it is constrained by the hadron density $\rho_h$ of the HRG model in hadronic temperature, whereas, for quark temperature range, beyond the transition temperature, jet transport coefficient at the initial time of QGP formation, $\hat{q}_0$ ($\approx 0.9$ GeV$^2$/fm) is used. One can find that the estimation of $\hat{q}/T^3$ in quasi-particle model (red dashed line) is comparable to the HRG estimation (solid blue line) in the hadronic temperature range. 
\par
The green dash-dotted and solid lines in Fig.~\ref{qbyT3_all} (left) show results based on pQCD + HTL approximation~\cite{pQCDHTL} for quark jets of initial energy 2 and 10 GeV respectively. It is interesting to observe that the results based on the pQCD + HTL approximation are quite comparable to that obtained from quasi-particle model based on Eq.~(\ref{qh_rho_qN}). Therefore, one may relate the estimation based on $\hat{q}_N$ in the quasi-particle model to a weakly coupled QGP system, as traditionally described by pQCD calculations. In Fig.~\ref{qbyT3_all} (right), the red dashed line shows $\hat{q}$ estimation using Eq.~(\ref{qh_rho_q0}) starting from high (QGP) to low (hadronic) temperature taking $\hat{q}$ at the initial time of QGP formation ($\hat{q}_0$ $\approx$ 0.9 GeV$^2$/fm) as reference point. Interestingly, $\hat{q}/T^3$ from Eq.~(\ref{qh_rho_q0}) shows a mild peak structure near the transition temperature. Analyzing $T$ dependence of Eq.~(\ref{qh_rho_qN}) and Eq.~(\ref{qh_rho_q0}), one can recognize $\hat{q}\propto g(T)\times T^3$ and $\hat{q}\propto g(T)$, respectively. Therefore, their normalizing function will be $\hat q/T^3$~$\propto g(T)$ and $\hat q/T^3$~$\propto$~$g(T)/T^3$, respectively. For the latter case, the peak structure is obtained as a result of the combined effects of increasing $g(T)$ and decreasing $1/T^3$.
\par
The magenta dash-dotted and solid lines in Fig.~\ref{qbyT3_all} (right) show semi-QGP + magnetic monopoles results~\cite{Gyulassy2} for quark jets of initial energy 2 and 10 GeV respectively. The cyan solid line presents the estimation of $\hat q/T^3$ for quark jets with momentum 10 GeV/$c$ at zero chemical potential using a dynamical quasiparticle model (DQPM)~\cite{DQPM}. The black lines in Fig.~\ref{qbyT3_all} (right) depict the jet transport coefficient in absence of magnetic field for an isotropic $\cal{N}$ = 4 SYM plasma of the form~\cite{Liu:2006ug}:
\begin{equation}
  \hat{q}_0 = \frac{\pi^{\frac{3}{2}} \Gamma(\frac{3}{4})}{\Gamma(\frac{5}{4})} \sqrt{\lambda}\,  T^3\,
  \label{q0_SYM}
\end{equation}
where $\sqrt{\lambda} = \sqrt{g_{YM}^2\, N_c}$ with $g_{YM}$ denoting the coupling strength of the strongly coupled plasma with color degeneracy factor $N_c$. The dash-dotted and solid black lines correspond to $\lambda$ = 4$\pi$ and 8$\pi$ respectively. It is observed that the semi-QGP + magnetic monopoles based results show qualitatively similar behaviour to that obtained from quasi-particle model based on Eq.~(\ref{qh_rho_q0}). Both have a peak structure near transition temperatures of their respective models and the DQPM calculation lies between these two predictions. One may, therefore, relate $\hat{q}_0$-based estimation in the quasi-particle model with a strongly coupled QGP system. 
\par
It is quite interesting to observe a two-directional aspect of QCD matter by calibrating with existing experimental knowledge of $\hat{q}$ for cold nuclear matter ($\hat{q}_N\approx 0.02$ GeV$^2$/fm) and for hot QGP ($\hat{q}_0\approx 0.9$ GeV$^2$/fm). When one approaches from the hadronic phase to the QGP phase with $\hat{q}_N$ as reference point, the results obtained are close to those of a weakly coupled QGP system. On the other hand, when one approaches from QGP to hadronic phase with $\hat{q}_0$ as reference point, the results favour a strongly coupled QGP system.

After calibrating the numerical estimations of the two possible simple (quasi-particle model based) expressions with existing values of weakly and strongly coupled QGP systems, the next aim is to see their changes due to finite magnetic field, which is discussed in the next subsection. The main focal point of present work is the relative changes in $\hat{q}$ due to magnetic field, therefore the absolute values of $\hat{q}$($B$ = 0) may be considered as a reference point only. 

\subsection{$\hat{q}$ in presence of finite magnetic field}
\begin{figure*}[h!]
  \centering
  \includegraphics[width=0.48\textwidth]{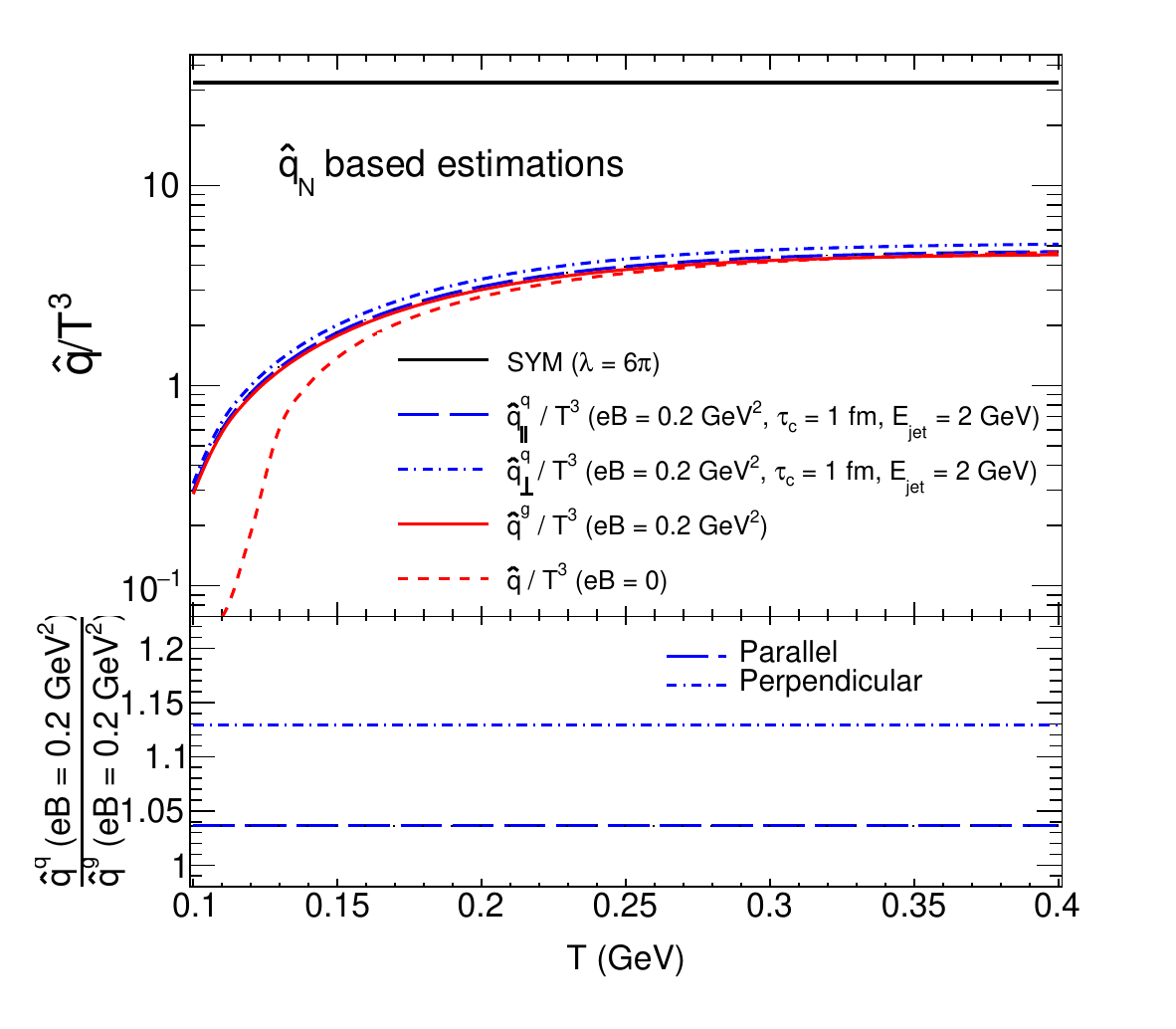}
  \includegraphics[width=0.48\textwidth]{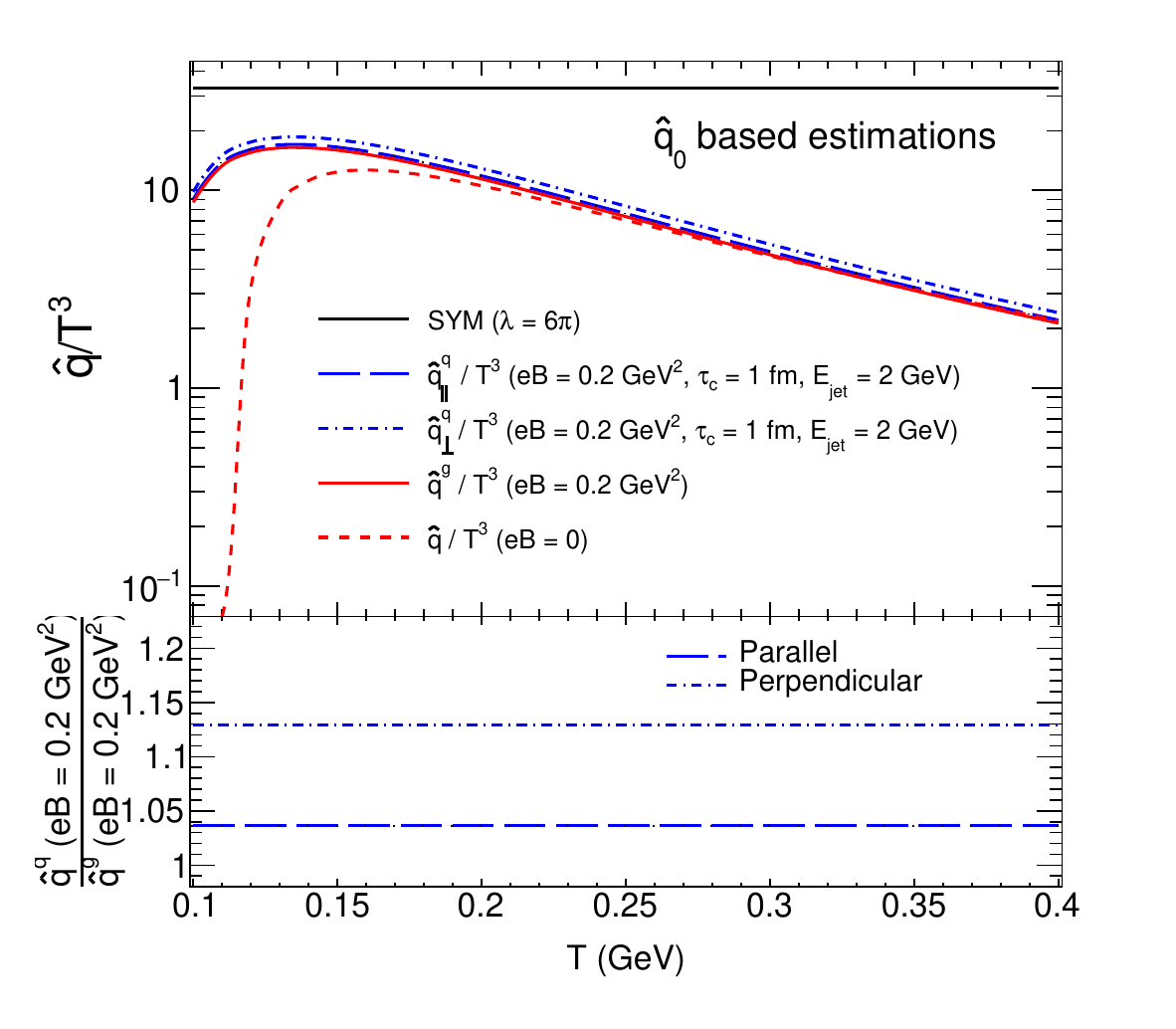}
  \caption{Temperature (T in GeV) dependence of $\hat{q}^q_{\parallel,\perp}/T^{3}$ (blue long dashed and dash-dotted lines) at $eB$ = 0.2 GeV$^{2}$, $\tau_{c}$ = 1 fm are depicted for quark jet of initial energy $E_{jet}$ = 2 GeV using Eqs.~(\ref{qh_q0}) \& (\ref{qh_rho_qN}) (left) and Eqs.~(\ref{qh_q0}) \& (\ref{qh_rho_q0}) (right). The red solid line shows T dependence of $\hat{q}^g/T^{3}$ for gluon jet at $eB$ = 0.2 GeV$^{2}$ obtained using Eq.~(\ref{qh_rho_qN}) (left) and Eq.~(\ref{qh_rho_q0}) (right). Red dashed lines of both figures represent $\hat{q}/T^{3}$ at $eB$ = 0 estimated using Eq.~(\ref{qh_rho_qN}) (left) and Eq.~(\ref{qh_rho_q0}) (right). Black solid line corresponds to an isotropic $\cal{N}$ = 4 SYM plasma at zero magnetic field for $\lambda$ = 6$\pi$ [Eq.~(\ref{q0_SYM})]. The ratios between $\hat{q}^{q}_{\parallel,\perp}$ and $\hat{q}^{g}$ in presence of magnetic field are shown by blue long dashed and dash-dotted lines in the bottom panels.}
  \label{s_eta_q0}
\end{figure*}

The temperature dependence of $\hat{q}/T^{3}$ in presence of magnetic field taking $\hat{q}_N$ as reference point as described by Eq.~(\ref{qh_rho_qN}) and taking $\hat{q}_0$ as reference point as described by Eq.~(\ref{qh_rho_q0}) are shown in left and right panels of Fig.~\ref{s_eta_q0}, respectively. 
The solid red lines in top panels of Fig.~\ref{s_eta_q0} show $\hat{q}/T^{3}$ for gluon jets ($\hat{q}^g$) at finite magnetic field, $eB$ = 0.2 GeV$^2$, estimated from Eq.~(\ref{q_gluon}).  The parallel and perpendicular components of $\hat{q}/T^{3}$ at $eB$ = 0.2 GeV$^2$ for quark jets of initial energy 2 GeV are shown by long dashed and dash-dotted blue lines, respectively. These results are obtained from the parallel ($\hat{q}^q_{\parallel}$) and perpendicular ($\hat{q}^q_\perp$) expressions, given in Eqs.~(\ref{q_par_quark}) and (\ref{q_per_quark}), respectively. The solid black lines depict the jet transport coefficient for an isotropic $\cal{N}$ = 4 SYM plasma at $\lambda$ = 6$\pi$. The dashed red lines representing $\hat{q}/T^{3}$ in absence of magnetic field are also shown in top panels of Fig.~\ref{s_eta_q0} for reference. 
\par
Qualitatively the temperature dependence of $\hat{q}/T^{3}$ in presence of magnetic field is found to have similar behaviour to that in absence of magnetic field. At lower temperature, significant enhancement of $\hat{q}/T^{3}$ is observed in presence of magnetic field and the enhancement reduces with the increasing temperature for both gluon and quark jets. If one analyzes the detailed anatomy of $\hat{q}/T^3$ of quark jet, given in Eqs.~(\ref{q_par_quark})-(\ref{q_per_quark}), then one can identify two sources of $B$-dependent components. One is the $B$-dependent degeneracy factor, for which jet transport coefficient gets enhanced, and the other is the phase space part made of $\tau_B$, for which $\hat{q}$ is further enhanced. However, for gluon jets, later component is missing due to their chargeless nature. The bottom panels of Fig.~\ref{s_eta_q0} show ratios of parallel and perpendicular components of $\hat{q}^q$ for quark jets to $\hat{q}^g$ for gluon jets in presence of magnetic field. It is interesting to observe that, in comparison to gluon jets, the parallel and perpendicular components of quark jets experience an increase in $\hat{q}$  values of around 4\% and 13\%, respectively due to presence of magnetic field. So, one can indirectly observe two interesting aspects of QGP at finite magnetic field via jet probe. Former part of quark jet contains magneto-thermodynamics of QCD matter via $g(T, B)$ and its later part is connected with thermodynamical phase  shrinking due to cyclotron motion of quark jet. Therefore, the present investigation is hinting at a possibility of difference between jet quenching phenomena for quarks and gluons in presence of magnetic field. 

The picture can be understood as follows. During the jet quenching mechanism, two components are involved. One is the probe, either quark-initiated or gluon-initiated jets (hereafter referred as quark jet and gluon jet respectively), which will eventually produce a collimated shower of particles. Another is the medium, which is assumed to be gluon dominated. Now, in presence of magnetic field, any thermodynamical quantity like gluon density will be modified. This modification is connected with the rich quark-condensate or constituent quark mass profile as a function of $T$ and $B$. Recent LQCD calculations~\cite{Bali1,Bali2} predicted an inverse magnetic catalysis profile near the transition temperature, for which the transition temperature decreases with increasing magnetic field. This LQCD-based $T$, $B$-dependent quark condensate profile will build magneto-thermodynamical phase space of different thermodynamical quantities like entropy density, pressure etc. We have tuned that profile via $T$, $B$-dependent degeneracy factor $g(T,B)$. Calculating the corresponding gluon-dominated medium density at finite $B$ using this $g(T,B)$, we have incorporated $B$-dependent information of medium into jet transport coefficient $\hat{q}$ as it is proportionally connected to the density of the medium.

Next, let us discuss the $B$-dependent information of the probe (quark and gluon jet). For gluon jet, no modification is possible but for quark jet, modification comes from the Lorentz force. A mild bending of quark jet propagation is expected due to its cyclotron motion, quantified through $\tau_B$. This $B$-dependent information has been indirectly incorporated in the present work through the $B$-dependent phase space factor of shear viscosity to entropy density ratio. 
\par
This impact of magnetic field on $\hat{q}$ might be challenging to observe experimentally, however, one can perform a comparative study by measuring nuclear suppression factor $R_{\rm AA}$ separately for quark and gluon jets in central (where $eB$ is expected to be zero) and non-central (where $eB$ is expected to be non-zero) heavy-ion collisions. The quarks being charged will be affected more than gluons in presence of magnetic field. Therefore, if the experiment can be designed for these two separate measurements, the difference between two measurements can be attributed to the $B$ dependence of jet transport coefficient. Recently measurements of directed flow ($v_1$) for hadrons by STAR~\cite{STAR_v1} and ALICE~\cite{ALICE_v1} collaborations are linked with the impact of electromagnetic field on the medium, inspired from theoretical prediction~\cite{SDas_v1}. In this direction, future measurements are also planned by ATLAS and CMS collaborations~\cite{EPJ_conf}.

\subsection{Comparison to AdS/CFT Correspondence}
\label{ads}
The jet transport coefficient $\hat{q}$ for the strongly coupled QCD plasma in presence of a magnetic field has also been computed directly using the non-perturbative toolbox of the anti–de Sitter/conformal field theory (AdS/CFT) correspondence~\cite{MALD,GUBS1,GUBS2,Li,Ma,Mamo,Rougemont}.
\begin{figure*}[h!]
  \centering
  \includegraphics[width=0.6\textwidth]{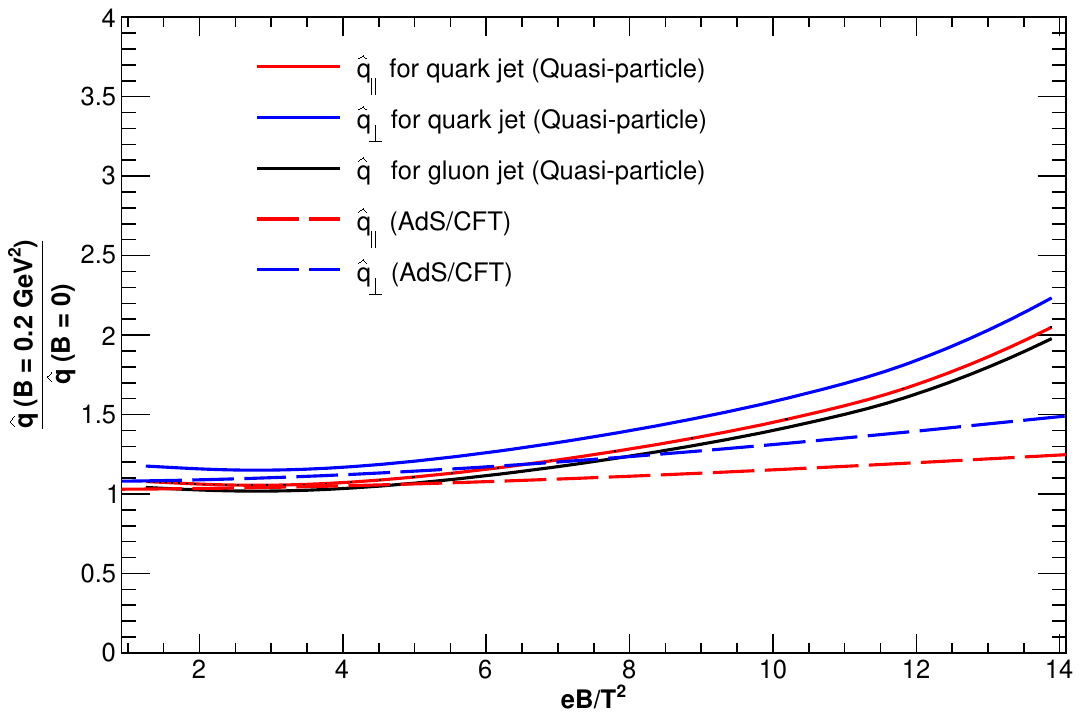}
  \caption{Comparison of the results for the ratio of $\hat{q}(B)$ to $\hat{q}(B=0)$ obtained from the quasi-particle like approach against AdS/CFT correspondence for $eB$ = 0.2 GeV$^{2}$ and $\tau_c$ = 1 fm. The results for the quasi-particle approach are shown as solid lines, and those from AdS/CFT correspondence~\cite{Rougemont} are shown by the dashed lines.}
  \label{fig:qh_with_AdS}
\end{figure*}
It has been shown that compared to the pQCD estimation, the magnitude of $\hat{q}(B=0)$ calculated via the AdS/CFT correspondence is closer to that extracted from RHIC data~\cite{ESKO,Dain}. If the magnetic field is along the $z$-direction, the parton that forms the jet while traversing through the medium can suffer a momentum broadening either along or transverse to the magnetic field direction, giving rise to three different components of jet transport coefficient: $\hat{q}_{\parallel(\perp)}$, $\hat{q}_{\perp(\parallel)}$ and $\hat{q}_{\perp(\perp)}$, where $\hat{q}_\parallel \equiv \hat{q}_{\parallel(\perp)}$ and $\hat{q}_\perp \equiv \hat{q}_{\perp(\parallel)} + \hat{q}_{\perp(\perp)}$. In this notation, the first and second symbols denote the direction of the moving parton and the direction in which the momentum broadening of the jet occurs with respect to the magnetic field direction, respectively. 
\par
Figure~\ref{fig:qh_with_AdS} shows an interesting comparison for the ratio $\hat{q}(B)/\hat{q}(B=0)$ obtained from quasi-particle model using Eq.~(\ref{q_par}) against the AdS/CFT correspondence results for values of $eB/T^2$ up to 14. Both find $\hat{q}_{\perp} > \hat{q}_{\parallel}$ and support enhancing trend of jet transport coefficient in presence of magnetic field, however, a quantitative difference is observed between them. One of the possible reasons might be that the current estimation of $\hat{q}$ from quasi-particle model has not adopted the quantum aspect of the magnetic field, where the phase space will be proportional to $eB$ due to Landau quantization. In that case, one might get a proportional $eB$ dependence as one grossly notice for the AdS/CFT correspondence results. In future, one can extend the present work in that direction.

\section{Summary}
\label{sum}
In summary, we presented the estimation of jet transport coefficient, $\hat{q}$ using a simple quasi-particle model where a temperature-dependent degeneracy factor $g(T)$ of partons is considered. The parameters of $g(T)$ are obtained by fitting the entropy density of lattice quantum chromodynamics. Although we are unable to properly fit other thermodynamical quantities, such as energy density and pressure, with the same set of parameters of $g(T)$, this model still provides a simple and easy-dealing tool which is consistent with QCD thermodynamics within a given range of temperature for the estimation of jet transport coefficient. Using the parametric degeneracy factor, we have estimated the temperature-dependent gluon density to obtain $\hat{q}(T)$ from hadronic to QGP phase by restricting low- and high-temperature values of $\hat{q}$ by their experimental values. During the restriction by low-temperature cold nuclear matter data, temperature-dependent $\hat{q}(T)$ from hadronic to QGP phase becomes quite close to the HTL or perturbative QCD results, which indicate transportation of jet in weakly interacting gas. On the other hand, $\hat{q}(T)$ estimation from QGP to hadronic phase is qualitatively similar to earlier estimate based on semi-QGP + magnetic monopoles, which corresponds to a strongly interacting liquid picture. 
\par
After calibrating our quasi-particle estimations with earlier results, we have gone through their finite magnetic field extensions. The effect of finite magnetic field is introduced in the model by replacing the temperature-dependent degeneracy factor $g(T)$ with temperature and magnetic field-dependent degeneracy factor  $g(T,B)$ whose parameters are obtained by fitting the magneto-thermodynamical data of lattice quantum chromodynamics. The jet transport coefficient $\hat{q} (T, B)$ is calculated for both quark and gluon jets. For quark jets, it splits into parallel and perpendicular components carrying magnetic field dependence from two sources: the field-dependent degeneracy factor and the phase space part guided from the shear viscosity to entropy density ratio. The value of $\hat{q}$ is found to be enhanced due to the collective role of these two sources for quark jets and, in case of gluon jets, only the field-dependent degeneracy factor comes into play. The $\hat{q}$ for gluon jets and both the parallel and perpendicular components for quark jets show a significant enhancement at low temperature which gradually decreases towards high temperature. A similar enhancement of jet transport coefficients at finite magnetic field is also observed in the AdS/CFT correspondence calculations. Quasi-particle model results can provide additional information towards our current understanding and phenomenology of jet quenching in presence of the magnetic field. 
\par
It would be very interesting to experimentally perform a comparative study by measuring nuclear suppression factor $R_{\rm AA}$ separately for quark and gluon jets in central (where $eB$ is expected to be zero) and non-central (where $eB$ is expected to be non-zero) heavy-ion collisions as quarks being electrically charged are expected to experience the effect of the magnetic field more compared to the electrically neutral gluons.

\appendix
\section{}
\label{apendix}
In terms of the Fermi-Dirac (FD) distribution function of quarks and the Bose-Einstein (BE) distribution function of gluons, the energy density ($\epsilon$) of the QGP system can be expressed as
\begin{align}
  \epsilon_{QGP} ={}& \frac{g_{g}}{(2\pi)^{3}} \int_{0}^{\infty} \frac{\omega_{g}}{e^{\beta\omega_{g}} - 1} {d^{3}\vec{k}}\nonumber\\
  & + \frac{g_{Q} }{(2\pi)^{3}}\int_{0}^{\infty} \frac{\omega_{Q}}{e^{\beta\omega_{Q}} + 1} {d^{3}\vec{k}}
  \label{e_QGP}
\end{align}
Here $\omega_{g}$ and $\omega_{Q}$ are energies and can be expressed as, $\omega_{g,Q} = \sqrt{\vec k^2+m_{g,Q}^2}$ and $\beta = 1/T$. Here $m_{g}$ and $m_{Q}$ are masses of quarks and gluons, however for massless QGP, $m_{g,Q} = 0$. Therefore for massless QGP, $\omega_{g,Q} = \vec k_{g,Q}$. If one convert the volume integral to line integral, $\int_{0}^{\infty} {d^{3}\vec{k}} \rightarrow 4\pi\int_{0}^{\infty}\vec{k}^{2} {d\vec{k}}$.\\
Eq.~\ref{e_QGP} can be expressed as,
\begin{align}
  \epsilon_{QGP} ={}& \frac{g_{g}}{(2\pi)^{3}} \int_{0}^{\infty} \frac{\vec{k_{g}}}{e^{\beta \vec{k_{g}}} - 1} {4\pi \vec{k_{g}}^{2} d\vec{k_{g}}}\nonumber\\
  & + \frac{g_{Q}}{(2\pi)^{3}} \int_{0}^{\infty} \frac{\vec{k_{Q}}}{e^{\beta \vec{k_{Q}}} + 1} {4\pi \vec{k_{Q}}^{2} d\vec{k_{Q}}}\nonumber\\
  ={}& \frac{g_{g}}{2\pi^{2}} \int_{0}^{\infty} \frac{\vec{k_{g}^{3}}}{e^{\vec{k_{g}}/T} - 1} {d\vec{k_{g}}}\nonumber\\
  & + \frac{g_{Q}}{2\pi^{2}} \int_{0}^{\infty} \frac{\vec{k_{Q}^{3}}}{e^{\vec{k_{Q}}/T} + 1} {d\vec{k_{Q}}}  \hspace{0.5cm} [\beta = 1/T]\nonumber\\
  ={}& \frac{g_{g}T^{4}}{2\pi^{2}} \int_{0}^{\infty} \frac{x^{3}}{e^{x} - 1} {dx}\nonumber\\
  & + \frac{g_{Q}T^{4}}{2\pi^{2}} \int_{0}^{\infty} \frac{y^{3}}{e^{y} + 1} {dy} \hspace{0.5cm} [\vec{k_{g}}/T = x, \vec{k_{Q}}/T = y]\nonumber\\
  ={}& \frac{g_{g}T^{4}}{2\pi^{2}} \int_{0}^{\infty} \frac{x^{3}}{e^{x} (1 - e^{-x})} {dx}\nonumber\\
  & + \frac{g_{Q}T^{4}}{2\pi^{2}} \int_{0}^{\infty} \frac{y^{3}}{e^{y} (1 + e^{-y})} {dy}\nonumber\\
  ={}& \frac{g_{g}T^{4}}{2\pi^{2}} \int_{0}^{\infty} x^{3}e^{-x} (1 - e^{-x})^{-1} {dx}\nonumber\\
  & + \frac{g_{Q}T^{4}}{2\pi^{2}} \int_{0}^{\infty} y^{3}e^{-y} (1 + e^{-y})^{-1} {dy}\nonumber\\
  ={}& \frac{g_{g}T^{4}}{2\pi^{2}} \int_{0}^{\infty} x^{3}e^{-x} \left[ \sum_{n = 0}^{\infty} e^{-nx} \right] {dx}\nonumber\\
  & + \frac{g_{Q}T^{4}}{2\pi^{2}} \int_{0}^{\infty} y^{3}e^{-y} \left[ \sum_{n = 0}^{\infty} (-1)^{n}e^{-ny} \right]{dy}\nonumber\\
  ={}& \frac{g_{g}T^{4}}{2\pi^{2}} \sum_{n = 0}^{\infty} \int_{0}^{\infty} x^{3}e^{-(1+n)x} {dx}\nonumber\\
  & + \frac{g_{Q}T^{4}}{2\pi^{2}} \sum_{n = 0}^{\infty} (-1)^{n} \int_{0}^{\infty} y^{3}e^{-(1+n)y} {dy}
  \label{e_QGP1}
\end{align}
If one consider, (1+n)x = a and (1+n)y = b then, Eq.~\ref{e_QGP1} can be represented as,
\begin{align}
  \epsilon_{QGP} ={}& \frac{g_{g}T^{4}}{2\pi^{2}} \sum_{n = 0}^{\infty} \frac{1}{(n+1)^{4}} \int_{0}^{\infty} a^{3}e^{-a} {da}\nonumber\\
  & + \frac{g_{Q}T^{4}}{2\pi^{2}} \sum_{n = 0}^{\infty} (-1)^{n} \frac{1}{(n+1)^{4}} \int_{0}^{\infty} b^{3}e^{-b} {db}
  \label{e_QGP2}
\end{align}
Simplification of $\int_{0}^{\infty} t^{3}e^{-t} {dt}  = \Gamma (4) = 6$ and expanding binomially one can get, $\sum_{n = 0}^{\infty} \frac{1}{(n+1)^{4}} = \xi(4) = \frac{\pi^{4}}{90}$ and $\sum_{n = 0}^{\infty} (-1)^{n} \frac{1}{(n+1)^{4}} = \frac{7}{8} \xi(4) =\frac{7}{8} \frac{\pi^{4}}{90} $. 
\begin{align}
  \epsilon_{QGP} &= \frac{g_{g}T^{4}}{2\pi^{2}} \frac{6\pi^{4}}{90}+ \frac{g_{Q}T^{4}}{2\pi^{2}}\frac{7}{8} \frac{6\pi^{4}}{90}\nonumber\\
  &= \Big[g_g+g_{Q}\Big(\frac{7}{8}\Big)\Big]\frac{3\pi^2}{90}T^4\approx 15.6~T^4
  \label{e_QGP3}
\end{align}
Pressure (P) of the QGP system follows the similar prescription as the energy density ($\epsilon$).\\

\section*{Acknowledgement}
The authors acknowledge following members of  \href{https://sabya623355158.wordpress.com/}{TPRC-IITBH} who previously worked on quasi-particle picture~\cite{SS_JPG,JD2}: Sarthak Satapathy, Jayanta Dey, Anki Anand, Ranjesh Kumar, Ankita Mishra and Prasant Murmu. D. Banerjee acknowledges the Inspire Fellowship research grant [DST/INSPIRE Fellowship/2018/IF180285]. A. Modak and P. Das acknowledge the Institutional Fellowship research grant of Bose Institute. Significant part of computation for this work was carried out using the computing server facility at CAPSS, Bose Institute, Kolkata.


\end{document}